\documentclass[journal]{IEEEtran}

\usepackage{amsmath,amssymb}
\usepackage{algorithm}
\usepackage{algpseudocode}
\usepackage{booktabs}
\usepackage{microtype}
\usepackage{multirow}
\usepackage{array}
\newcolumntype{L}[1]{>{\raggedright\arraybackslash}p{#1}}
\usepackage{tabularx}
\usepackage{graphicx}
\usepackage{xurl}
\usepackage[hidelinks]{hyperref}
\hypersetup{
  pdftitle={ExecWeave: Cross-Layer Execution Provenance for Observable Agentic Software},
  pdfsubject={Cross-layer attribution, runtime evidence, and empirical evaluation},
  pdfkeywords={AI agents, execution provenance, observability, runtime tracing}
}
\usepackage{tikz}
\usepackage{xspace}
\usetikzlibrary{arrows.meta,positioning,fit,calc}

\newcommand{\system}{\textup{\textsc{ExecWeave}}\xspace}
\newcommand{\etal}{et al.}

\newcommand{\notnative}{Not documented}

\title{ExecWeave: Cross-Layer Execution Provenance for Observable Agentic Software}

\author{Guo-Wei~Wong, and Meng~Chang~Chen%
\thanks{Guo-Wei Wong is with National Taiwan University, Taipei, Taiwan.}%
\thanks{Meng Chang Chen is with Academia Sinica, Taipei, Taiwan.}%
}

\begin{document}
\maketitle

\begin{abstract}
Agentic software invokes tools, launches processes, accesses files, and communicates with local or remote services. Application traces describe logical operations, whereas operating-system telemetry records machine activity without necessarily identifying the responsible agent action. Relating these evidence planes is a reconstruction problem: a plausible but unsupported association can assign an effect to the wrong principal.

We present \system, a local-first system that materializes provider semantics and independently observed runtime events into a typed execution graph. The implementation retains separate source artifacts, distinguishes logical occurrences from repeated evidence, and records attribution, inference, and causal-status metadata. An optional command-to-process correlator emits a non-causal inferred bridge only when its bounded search yields one compatible observed process; uniqueness within this observed candidate set is not a guarantee of causal correctness. OS-observed file and network effects acquire a semantic tool or agent principal only through an observed process plus a justified command-to-process association; session-correlated file-system events do not by themselves provide that principal.

We evaluate the frozen v0.8.25 implementation on Linux using direct portable-collector sampling measurements, fixed-work overhead measurements, scoped graph-materialization benchmarks, one real AutoGen--Ollama integration run, and targeted negative controls for the optional correlator. The real integration validates end-to-end semantic capture, independent runtime observation, process-reference resolution, and graph materialization. The negative controls confirm abstention under observed ambiguity while exposing false uniqueness when evidence is missing. Direct RuntimeCollector measurements expose the expected sampling blind window, while fixed-work trials show substantial portable-observation cost: median paired elapsed-time overhead ranges from 62.3\% to 408.8\% across the three tested workloads. Cross-OS parity, broad provider coverage, product superiority, and human debugging benefit remain unmeasured. Source code is available at \url{https://github.com/Irish-kw/ExecWeave}.
\end{abstract}


\section{Introduction}
\IEEEPARstart{A}{gentic} software is changing the unit of execution that developers must understand. A conventional application can often be inspected through instrumented function and service boundaries. An agentic application may additionally select tools dynamically, delegate work to other agents, issue shell commands, start language runtimes, call local model servers, edit repositories, and contact external endpoints as part of a single user request. The externally visible answer is therefore only the end of a longer execution chain whose intermediate effects may matter for debugging, security, reproducibility, and accountability.

The observability ecosystem has responded rapidly. LangSmith records traces composed of runs and integrates with agent frameworks and OpenTelemetry~\cite{langsmith2026observability,langsmith2026otel}. Langfuse records observations and renders agent graphs inferred from typed observations and nesting~\cite{langfuse2026agentgraphs,langfuse2026tracing}. Phoenix uses OpenTelemetry and OpenInference to trace model calls, tools, retrieval, and custom application logic~\cite{phoenix2026tracing,phoenix2026otel}. AgentOps organizes agents, language-model calls, actions, and tools into sessions and hierarchical spans~\cite{agentops2026,agentops2026core}, while W\&B Weave records operations and calls and provides trace-tree, flame, composition, and graph-oriented views~\cite{wandb2026tracing,wandb2026traceview}. Academic systems similarly convert agent execution into analyzable graph structure. AgentGraph transforms traces into knowledge graphs with agents, tasks, tools, data, and human interventions~\cite{wu2026agentgraph}; AgentSeer separates temporal action and component graphs~\cite{wicaksono2026agentseer}; Graph of Trace records scientific-agent tool calls and code execution as an incrementally updated directed graph~\cite{gao2026graphtrace}; and AlSayyad \etal's AgentTrace proposes structured operational, cognitive, and contextual telemetry for agent systems~\cite{alsayyad2026agenttrace}.

These systems establish graph-oriented workflow inspection as an important observability capability. Relating that workflow to independent machine observations raises a distinct reconstruction problem. Agent-level instrumentation describes what an integration point \emph{reported}: an agent emitted a tool call, a framework entered a function, or a model returned a response. Operating-system observation describes what a machine \emph{observed}: a process was created, a file changed, or a socket appeared. The two descriptions are related but not equivalent. A declared shell command does not prove that the corresponding process executed; a file body supplied by a provider hook does not prove an operating-system read; and a process connecting to a network endpoint does not by itself identify which logical agent action requested that connection.

We call this separation the \emph{semantic--runtime observability gap}. It has two consequences. First, semantic-only traces can omit effects below the instrumentation boundary. Second, OS-only telemetry can preserve effects while losing the logical principal or tool invocation that motivated them. Naively joining the two streams by timestamp is unsafe under concurrency, process reuse, wrappers, shell indirection, shared runtimes, or overlapping sub-agent activity. For accountability-oriented observability, a false attribution can be more damaging than an explicit unknown: it can blame Agent A for an effect caused by Agent B.

\system addresses this problem by treating observations as evidence and the graph as a materialized, queryable view of that evidence. Provider/framework semantics, inference-gateway observations, model-runtime observations, and OS-runtime evidence remain distinguishable after graph construction. Command-correlated bridges are marked as inferred and non-causal. Multiple compatible observed candidates produce no bridge rather than a score-based choice. This design follows a lineage of provenance systems that emphasize faithful derivation histories~\cite{muniswamy2006pass,pohly2012hifi,bates2015lpm,pasquier2017camflow}, while specializing the evidence model for agent runtimes whose relevant identities include agents, sub-agents, tool calls, model requests, provider sessions, processes, files, and network endpoints.

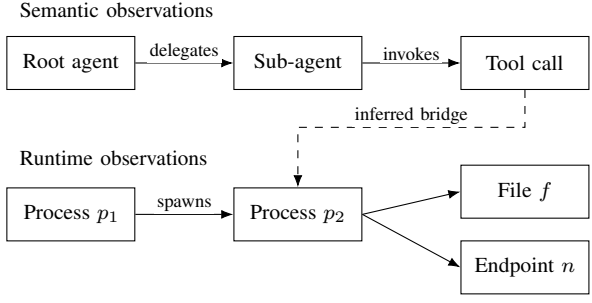
\begin{figure}[t]
\centering
\begin{tikzpicture}[
 b/.style={draw,align=center,text width=1.55cm,minimum height=7mm,inner sep=2pt,font=\footnotesize},
 e/.style={-{Latex[length=1.6mm]},thin},
 l/.style={font=\scriptsize,fill=white,inner sep=1pt}
]
\node[b] (a) at (0,0) {Root agent};
\node[b] (b) at (3,0) {Sub-agent};
\node[b] (t) at (6,0) {Tool call};
\node[anchor=west,font=\footnotesize] at (-0.8,0.7) {Semantic observations};
\node[b] (p) at (0,-2) {Process $p_1$};
\node[b] (q) at (3,-2) {Process $p_2$};
\node[b] (f) at (6,-1.7) {File $f$};
\node[b] (n) at (6,-2.7) {Endpoint $n$};
\node[anchor=west,font=\footnotesize] at (-0.8,-1.25) {Runtime observations};
\draw[e] (a) -- node[l,above]{delegates} (b);
\draw[e] (b) -- node[l,above]{invokes} (t);
\draw[e] (p) -- node[l,above]{spawns} (q);
\draw[e] (q.east) -- (f.west);
\draw[e] (q.east) -- (n.west);
\draw[e,dashed] (t.south) -- (6,-0.85) -- node[l,above]{inferred bridge} (3,-0.85) -- (q.north);
\end{tikzpicture}
\caption{Conceptual semantic--runtime gap, not a measured run. Solid edges belong to one observation plane and retain that plane's provenance; process-attributed file edges require a suitable backend. A dashed cross-layer bridge is a separate inferred claim, not causal proof.}
\label{fig:gap}
\end{figure}

Figure~\ref{fig:gap} illustrates the central distinction. The semantic plane can say that a root agent delegated to a sub-agent and that the sub-agent invoked a tool. The runtime plane can say that a process spawned a child that modified a file and contacted a service. The research problem is not to draw both planes; it is to determine which cross-plane relations are justified, with what evidence, and what should remain unknown.

This paper makes four contributions.
\begin{itemize}
  \item \textbf{An evidence-typed execution-graph contract for agentic software.} We define a typed graph that preserves agent, tool, model, process, file, network, and runtime identities together with evidence source, causality, attribution method, and temporal support. The contribution is the explicit agent-oriented evidence contract, not the general idea of combining application and OS observations.
  \item \textbf{Conservative command-to-process attribution with explicit limits.} We formulate that link as a partial mapping in which abstention is a first-class outcome. The optional correlator searches a bounded window with executable and command-line heuristics and emits a bridge only when the observed candidate set is a singleton. The bridge is marked inferred and non-causal. Uniqueness inside the observed set is not causal identification and is not a file- or socket-level claim.
  \item \textbf{Evidence-preserving live materialization across heterogeneous integrations.} \system separates raw runtime, semantic, and correlated artifacts; normalizes provider-specific events into a canonical model; supports portable observation on Windows, macOS, and Linux together with a stronger Linux syscall-oriented reference backend; and uses the same graph contract for live and finished inspection. These are implementation capabilities; the empirical campaign in this paper is Linux-scoped.
  \item \textbf{A scoped empirical validation of evidence fidelity and system cost.} We validate one real AutoGen--Ollama execution path, directly measure portable-collector blind windows, quantify fixed-work overhead, and characterize graph materialization and serialization. Targeted attribution negative controls are retained only to bound the optional heuristic and its abstention behavior, not as a primary accuracy benchmark.
\end{itemize}

The remainder of the paper first motivates the problem and separates it from adjacent tracing and provenance work, then formalizes the evidence model and describes \system's design and implementation. We next present the measured evaluation methodology and results, followed by limitations, threats to validity, and implications for future agent observability standards.

\section{Background and Motivation}
\subsection{Three views of one agent execution}
Consider a coding agent that receives a request to modify a repository. The root agent delegates test diagnosis to a sub-agent. The sub-agent invokes a shell tool, which starts a command interpreter and then a language runtime. The runtime reads several source files, writes a report, and contacts a local or remote model endpoint. Three observers can produce three different records of this run.

An \emph{agent-semantic observer} can record the user prompt, agent identity, delegation, model request, tool name, tool arguments, tool result, and final response. This record is close to the logical workflow and is invaluable for understanding model decisions. Its boundary, however, is the provider or framework integration point. If the tool internally launches another process, touches an undeclared file, or uses a library that performs a network request, these effects need not appear as first-class semantic events.

A \emph{runtime observer} can instead record the process tree, file accesses, filesystem mutations, and network connections. Kernel-level provenance systems have long shown the value of such dependency histories for forensics and information-flow analysis~\cite{pohly2012hifi,bates2015lpm,pasquier2018runtime}. Runtime evidence is closer to machine effects but can lack application semantics. A process identifier does not reveal whether it corresponds to a root agent, a sub-agent, a tool, a model runtime, or unrelated work that happens to overlap in time.

A \emph{cross-layer observer} attempts to preserve both views and the evidence that justifies relations between them. This creates a qualitatively different requirement: the system must avoid collapsing ``reported by the agent'' and ``observed by the machine'' into the same truth category. It must also represent uncertainty explicitly. \system is designed around this third view.

\subsection{Why distributed tracing is necessary but insufficient}
Distributed tracing established the practical value of propagating request context across software boundaries. Dapper showed that low-overhead tracing can reconstruct request paths through large distributed services~\cite{sigelman2010dapper}, while X-Trace pursued cross-layer diagnosis by propagating metadata across protocols and components~\cite{fonseca2007xtrace}. OpenTelemetry defines APIs, semantic conventions, context propagation, and a span-based trace model~\cite{otel2026}.

Agent platforms appropriately build on this ecosystem. OpenTelemetry-compatible spans are a strong representation for instrumented operations and can carry arbitrary attributes. The limitation is not expressive power: a sufficiently instrumented application can place almost any fact in a span. The limitation is \emph{observation independence}. If a tool or process is not instrumented, or if the semantic trace records a declaration rather than an OS effect, application spans do not independently establish that the effect occurred. Conversely, an OS observer can see an effect without a propagated span context. \system therefore treats OpenTelemetry-style application tracing as complementary semantic evidence rather than a substitute for independent runtime observation.

\subsection{Why provenance alone is also insufficient}
Data and whole-system provenance systems model derivation and causality as graphs. PASS integrated provenance with storage~\cite{muniswamy2006pass}; Hi-Fi captured high-fidelity whole-system provenance in the Linux kernel~\cite{pohly2012hifi}; SPADE separated provenance collection, storage, and querying~\cite{gehani2012spade}; Linux Provenance Modules emphasized trustworthy whole-system provenance~\cite{bates2015lpm}; CamFlow provided practical, configurable whole-system provenance capture~\cite{pasquier2017camflow}; and later work used streaming provenance for runtime security analysis~\cite{pasquier2018runtime}. ProTracer and RAIN further explored the trade-off between collection overhead and causal precision~\cite{ma2016protracer,ji2017rain}.

These systems motivate \system's evidence discipline, but agentic software introduces semantic entities that are not reducible to OS objects. A sub-agent may share a process with its parent. A tool call may execute through a provider-managed runtime. Two logical agents may issue commands through the same executable. A remote model invocation may be semantically visible while the remote process is, correctly, outside local OS visibility. Therefore, a useful agent execution graph must represent both logical and runtime identities without pretending that one uniquely determines the other.

\subsection{Design objectives}
From these observations we derive five objectives.

\textbf{O1: Evidence separation.} Direct runtime observations, provider-supplied semantics, exact shared identities, and heuristic correlations must remain distinguishable after normalization and graph materialization.

\textbf{O2: Conservative attribution.} If several runtime entities are consistent with a semantic action, the system should preserve the ambiguity rather than select the visually nearest candidate.

\textbf{O3: Provider and OS neutrality at the graph boundary.} Integration-specific telemetry should map into a stable graph vocabulary so that downstream analysis does not require one visualization implementation per provider or operating system.

\textbf{O4: Incremental consistency.} Completed-live and standalone inspection should converge on the same finalized graph and projection policy. Provisional incremental state and display-only links must remain distinguishable from archived evidence.

\textbf{O5: Explicit fidelity limits.} Sampling blind windows, unsupported provider internals, remote execution boundaries, and non-causal filesystem observations must be represented as limitations rather than converted into negative evidence.

\section{Related Work}
\subsection{Agent observability and trace visualization}
Contemporary agent observability products focus on instrumented application behavior. LangSmith structures a trace as runs and supports framework integrations, manual instrumentation, and OpenTelemetry ingestion~\cite{langsmith2026observability,langsmith2026otel}. Langfuse records typed observations and can render agent graphs either from agentic observation types and nesting or from LangGraph integrations~\cite{langfuse2026agentgraphs}. Phoenix is built around OpenTelemetry/OpenInference instrumentation and traces model calls, retrieval, tools, and custom application logic~\cite{phoenix2026tracing}. AgentOps uses OpenTelemetry-based hierarchical spans and session views for LLM calls, actions, tools, errors, timing, and cost~\cite{agentops2026core}. W\&B Weave records operations and calls with parent--child relationships and exposes trace-tree, flame, composition, and graph-oriented views~\cite{wandb2026tracing,wandb2026traceview}. Its MCP integration illustrates a relevant limitation of application-level tracing: client- and server-side operations can be separately observable without providing end-to-end visibility unless context is propagated~\cite{wandb2026mcp}.

Academic systems increasingly add explicit graph representations and structured trace models. AgentGraph converts execution logs to knowledge graphs whose nodes include agents, tasks, tools, data inputs/outputs, and humans, with typed edges linked to source trace spans; it additionally supports robustness and causal analyses~\cite{wu2026agentgraph}. AgentSeer constructs a temporal action graph and a component graph to expose action sequences and architectural relations~\cite{wicaksono2026agentseer}. Graph of Trace records scientific-agent tool calls and code execution and incrementally renders a directed workflow graph~\cite{gao2026graphtrace}. AlSayyad \etal's AgentTrace proposes a structured telemetry framework spanning operational, cognitive, and contextual surfaces~\cite{alsayyad2026agenttrace}. Wang \etal's 2026 survey organizes evidence tracing and execution provenance for LLM agents around a broader provenance taxonomy spanning retrieval, tools, memory, observability, auditing, and failure diagnosis~\cite{wang2026tracestotrust}. A separately named AgentTrace system reconstructs causal graphs from execution logs for post-hoc root-cause ranking~\cite{wang2026agenttracerca}; as described, its causal graph is log-derived rather than an independently collected OS evidence plane.

AgentSight is the closest prior system to the semantic--runtime problem studied here. It uses boundary tracing with eBPF to observe LLM traffic and kernel-level effects from outside the agent application, then correlates those streams across process boundaries~\cite{zheng2025agentsight}. Consequently, \system does \emph{not} claim novelty from combining agent semantics with OS observations in general. The distinction pursued here is narrower: an explicit evidence-typed artifact contract across heterogeneous provider/runtime adapters, non-causal labeling for heuristic cross-plane bridges, and an abstaining singleton policy whose supporting events remain inspectable. AgentSight's published design is Linux/eBPF-centered, whereas \system additionally exposes a lower-fidelity portable collector; the Linux empirical campaign reported here does not establish cross-OS superiority or parity.

\system likewise does not claim novelty from rendering agent traces as graphs. Its object is a graph whose nodes may originate from different observation planes and whose cross-plane edges retain evidence-source, inference, and causal-status metadata. The optional heuristic studied in this paper is specifically a command-to-process bridge; it is weaker than propagated identity and must remain absent when the observed candidate set is ambiguous.

\begin{table*}[t]
\centering
\caption{Positioning against representative agent-observability systems (sources checked September 2026). ``Not documented'' means the cited source does not describe independent OS process/file/network collection and agent-to-OS attribution as a native/default mechanism; it does not exclude custom ingestion of OS-derived spans.}
\label{tab:systems}
\footnotesize
\setlength{\tabcolsep}{3.0pt}
\renewcommand{\arraystretch}{1.05}
\begin{tabularx}{\textwidth}{@{}L{1.95cm}
>{\hsize=1.10\hsize\linewidth=\hsize\raggedright\arraybackslash}X
>{\hsize=0.85\hsize\linewidth=\hsize\raggedright\arraybackslash}X
>{\hsize=0.95\hsize\linewidth=\hsize\raggedright\arraybackslash}X
>{\hsize=0.95\hsize\linewidth=\hsize\raggedright\arraybackslash}X
>{\hsize=1.15\hsize\linewidth=\hsize\raggedright\arraybackslash}X@{}}
\toprule
System & Agent structure & Independent OS observation & Process/file/network evidence & Semantic$\rightarrow$runtime attribution & Documented focus \\
\midrule
LangSmith~\cite{langsmith2026observability,langsmith2026otel} & Nested runs and traces & \notnative & Custom-span dependent & User-instrumented & Application tracing, evaluation, and monitoring \\
Langfuse~\cite{langfuse2026agentgraphs,langfuse2026tracing} & Typed observations and agent graph & \notnative & Custom-span dependent & User-instrumented & LLM/agent observability and workflow visualization \\
Phoenix~\cite{phoenix2026tracing,phoenix2026otel} & OpenInference/OTel trace hierarchy & \notnative & Custom-span dependent & User-instrumented & AI tracing, evaluation, and experimentation \\
AgentOps~\cite{agentops2026,agentops2026core} & Session and hierarchical spans & \notnative & Custom-span dependent & User-instrumented & Agent session analytics and monitoring \\
W\&B Weave~\cite{wandb2026tracing,wandb2026traceview} & Ops/calls, tree and graph views & \notnative & Custom-span dependent & User-instrumented & Trace analysis integrated with evaluation \\
AgentGraph~\cite{wu2026agentgraph} & Trace-derived knowledge graph & \notnative & Input-trace dependent & Trace-derived & Interactive graph analysis and robustness testing \\
AgentSeer~\cite{wicaksono2026agentseer} & Temporal and component graphs & \notnative & Input-trace dependent & Trace-derived & Temporal action and architecture analysis \\
Graph of Trace~\cite{gao2026graphtrace} & Incremental directed execution graph & \notnative & Tool/code-event dependent & Event-derived & Scientific-agent workflow interpretation \\
AgentSight~\cite{zheng2025agentsight} & LLM/agent boundary semantics & Linux eBPF kernel observation & Process/file/system effects & Real-time cross-boundary correlation & System-level agent observability \\
\system & Unified typed execution graph & Portable OS collector; stronger Linux backend & Collected, backend-dependent & Optional singleton correlation, explicitly inferred/non-causal & Evidence-typed graph and inspectable cross-plane support \\
\bottomrule
\end{tabularx}
\end{table*}

\subsection{Distributed tracing and context propagation}
Dapper~\cite{sigelman2010dapper}, X-Trace~\cite{fonseca2007xtrace}, and OpenTelemetry~\cite{otel2026} motivate end-to-end identifiers and parent--child operation structure. \system borrows the principle that identities should be propagated when available, but does not assume that propagation is universal. Exact shared request identifiers can establish identity across integration points; in their absence, runtime correlation is explicitly weaker. This distinction is particularly important when a provider session ID, tool-call ID, or model request ID is a logical identifier rather than an operating-system PID.

\subsection{Whole-system provenance and attack investigation}
Whole-system provenance captures dependencies among processes and resources at a level below typical agent instrumentation. The W3C PROV model provides a general vocabulary of entities, activities, agents, and derivations~\cite{prov2013}; PASS, Hi-Fi, SPADE, LPM, and CamFlow demonstrate system-level provenance capture and querying~\cite{muniswamy2006pass,pohly2012hifi,gehani2012spade,bates2015lpm,pasquier2017camflow}. Runtime analysis over provenance graphs and attack-investigation systems such as ProTracer and RAIN show how causal reconstruction supports security tasks~\cite{pasquier2018runtime,ma2016protracer,ji2017rain}. On portable platforms \system has a weaker collection boundary than complete kernel provenance monitors. Its specialization is instead the explicit representation of agent, tool, model, conversation, provider, and runtime identities alongside OS entities and the evidence status of relations between them; application-level enrichment itself is not unique to \system.

\subsection{Execution-graph visualization}
Layered graph drawing has a long history. Sugiyama \etal~\cite{sugiyama1981hierarchical} formalized layered representations that reduce crossings and improve hierarchy readability; Gansner \etal~\cite{gansner1993directed} developed a practical multi-pass directed-graph drawing technique; and Brandes and K{\"o}pf~\cite{brandes2002horizontal} proposed efficient horizontal coordinate assignment. Dynamic visualization adds the problem of stability across graph updates. Human-subject studies show that mental-map preservation is task dependent rather than universally beneficial~\cite{archambault2011dynamic,archambault2013mentalmap}. Accordingly, \system treats layout as a presentation optimization constrained by evidence topology. Layout may route, bundle, compact, fold, or de-emphasize visual elements, but it must not mutate the source evidence graph. Display-only contracted paths require explicit provenance and non-causal labels.

\section{Problem Formulation}
\subsection{Evidence channels and canonical events}
Let an execution produce observations from a set of channels
\begin{equation}
\mathcal{C}=\{C_{sem},C_{gw},C_{model},C_{os}\},
\end{equation}
representing provider/framework semantics, inference-gateway evidence, model-runtime evidence, and operating-system evidence. A channel need not be available for every run. For exposition, write a normalized observation as
\begin{equation}
e=(id,s,t,\tau,r,a,\kappa),
\end{equation}
where $id$ is an event identity, $s$ and $t$ are optional source and target entities, $\tau$ is a timestamp, $r$ is a typed relation, $a$ contains attributes, and $\kappa$ denotes provenance metadata. This tuple is an abstraction, not the serialized schema: the implementation also carries a schema version, session ID, event type, and optional sequence number, and stores provenance fields within event attributes.

The vocabulary is extensible. Representative entity classes include
\begin{align}
V = {} & V_{agent}\cup V_{session}\cup V_{tool}\cup V_{command}\cup V_{model} \nonumber\\
 & \cup V_{runtime}\cup V_{process}\cup V_{executable} \nonumber\\
 & \cup V_{file}\cup V_{net}\cup V_{provider}\cup V_{other}.
\end{align}
The remaining types include directories, messages, tool-call occurrences, and unresolved process references. Canonical IDs, not display names, identify entities within a run; normalization must therefore supply appropriately scoped IDs.

\subsection{Materialized execution graph}
Given an ordered, schema-valid event prefix $E_{1:k}=\langle e_1,\ldots,e_k\rangle$, graph materialization produces
\begin{equation}
G_k=M(E_{1:k})=(V_k,A_k).
\end{equation}
Repeated observations of the directed tuple $(u,r,v)$ contribute to one edge. The implementation distinguishes a \emph{logical occurrence count} from a \emph{raw evidence count}. Semantic replay is folded only when an occurrence-shaped entity or an explicit occurrence discriminator supplies a replay key. Equal endpoints alone do not deduplicate repeated model calls. Supporting event IDs, backend sets, attribution modes, and first/last observation times are retained in the full graph; the live accumulator may omit per-edge event-ID lists to reduce memory use. Source-only or target-only lifecycle events update node metadata without manufacturing a dummy edge.

An entity-relation graph is consequently not an event-per-node graph. Seventeen separately recorded file opens may share one edge with a logical count of seventeen, whereas seventeen replays of one identified semantic occurrence can have logical count one and evidence count seventeen. Both differ from a claim that all real-world occurrences were captured. Node display attributes use first-available values, so a fixed input order is required for reproducible attribute selection when observations disagree.

\subsection{Cross-layer attribution as a partial mapping}
The command correlator addresses a restricted attribution task. Let $S$ be the command-declaration occurrences that it examines and $R$ the observed process identities. A declaration belongs to a tool-call entity, but is not necessarily in one-to-one correspondence with that entity. Define
\begin{equation}
\phi:S\rightarrow R\cup\{\bot\},
\end{equation}
where $\bot$ denotes abstention. For an eligible declaration $s$, let $K(s)$ contain the distinct observed processes passing the implementation's command-compatibility predicates within its time window. The implemented decision rule is
\begin{equation}
\phi(s)=
\begin{cases}
r, & K(s)=\{r\},\\
\bot, & \text{otherwise}.
\end{cases}
\label{eq:phi}
\end{equation}
Unsupported declarations also yield $\bot$. A heuristic score annotates an emitted match; it is neither a probability nor an acceptance threshold and does not resolve multiple compatible candidates. The bridge is attached to the declaration's tool-call entity and is marked inferred and non-causal. Uniqueness is local to the observed candidate set, not proof that unobserved processes do not exist, not a global one-to-one assignment, and not a causal-identification theorem.

\subsection{Evidence strength is not behavior severity}
\system assigns evidence grades to graph-supported findings. Table~\ref{tab:grades} summarizes their interpretation. The classifier first assigns D to an explicitly inferred edge or one recording an inference method; otherwise an explicitly non-causal edge receives C. Remaining edges are classified from their attribution modes and causal flag, taking the weakest applicable grade. A finding receives the weakest grade of its cited supporting edges in the order A, B, C, D, U; missing support gives U. Thus a heuristic bridge marked non-causal is D, not C. Mixed provenance does not unconditionally imply U because these precedence rules apply first.

\begin{table}[t]
\centering
\caption{Evidence grades, subject to the precedence rules in the text. These classify recorded provenance, not the trustworthiness of its emitter.}
\label{tab:grades}
\footnotesize
\setlength{\tabcolsep}{3.5pt}
\renewcommand{\arraystretch}{1.05}
\begin{tabularx}{\columnwidth}{@{}>{\centering\arraybackslash}p{0.75cm} X@{}}
\toprule
Grade & Meaning \\
\midrule
A & Recognized syscall attribution with an explicit causal flag. \\
B & Recognized sampled-process attribution with an explicit causal flag. \\
C & Session-observation attribution or explicitly non-causal support, unless classified as inference. \\
D & Explicit inference or recorded heuristic method. \\
U & Unclassified or insufficient attribution after the preceding rules, or missing supporting edges. \\
\bottomrule
\end{tabularx}
\end{table}

A severe behavior can have weak evidence and a benign behavior strong evidence. Grades are neither maliciousness scores nor calibrated probabilities. Their meaning depends on the collector and integration honoring the event contract; grading does not independently authenticate an event's claimed origin.

\subsection{Design invariants and their scope}
The following properties guide implementation and validation rather than asserting unrestricted guarantees.

\textbf{I1: No promotion of explicitly non-causal support.} An aggregate supported only by explicit non-causal flags remains non-causal. Conflicting explicit Boolean flags remain mixed. Missing flags are not equivalent to explicit negative flags; their treatment is specified below.

\textbf{I2: Observed-candidate abstention.} The command correlator emits no bridge when its eligible observed candidate set has zero or multiple process identities. This does not rule out false uniqueness caused by missing observations or coarse predicates.

\textbf{I3: Evidence preservation under presentation.} Layout and projection do not mutate the source graph. A contracted display path must remain distinguishable from a native evidence edge and retain its supporting path references.

\textbf{I4: Finalized-view convergence.} Completed-live and standalone views should agree when given the same finalized graph and projection settings. A pre-finalization incremental cache is not required to be byte-identical to that graph: it may omit support lists, use provisional references, and precede terminal semantic synchronization.

\section{System Design}
\subsection{Architecture}
\begin{figure*}[t]
\centering
\begin{tikzpicture}[
 b/.style={draw,rounded corners=1pt,align=center,text width=2.35cm,minimum height=9mm,inner sep=3pt,font=\footnotesize},
 e/.style={-{Latex[length=1.8mm]},thin},
 d/.style={-{Latex[length=1.8mm]},dashed,thin}
]
\node[b,fill=gray!10] (provider) at (0,1.55) {Agent / IDE / framework\\hooks and transcripts};
\node[b,fill=gray!10] (gateway) at (0,0) {Gateway / model-runtime\\integration};
\node[b,fill=gray!10] (os) at (0,-1.55) {OS collector\\process / file / network};
\node[b,text width=1.9cm] (sem) at (3.1,1.55) {Semantic\\sidecar};
\node[b,text width=1.9cm] (mid) at (3.1,0) {Gateway / model\\observations};
\node[b,text width=1.9cm] (raw) at (3.1,-1.55) {Runtime\\event stream};
\node[b,fill=gray!5] (norm) at (6.5,0) {Validation / normalization\\and stream assembly};
\node[b,fill=gray!5] (graph) at (10.5,0) {Canonical graph\\materialization};
\node[b,fill=gray!10] (view) at (13.8,0) {Live / finished views\\and graph queries};
\node[b,text width=3.3cm] (corr) at (8.6,-1.65) {Optional batch correlation\\complete stream plus inferred bridges};
\draw[e] (provider) -- (sem);
\draw[e] (gateway) -- (mid);
\draw[e] (os) -- (raw);
\draw[e] (sem.east) -- ++(0.55,0) |- (norm.north west);
\draw[e] (mid) -- (norm);
\draw[e] (raw.east) -- ++(0.55,0) |- (norm.south west);
\draw[e] (norm) -- (graph);
\draw[e] (graph) -- (view);
\draw[d] (norm.south) |- (corr.west);
\draw[d] (corr.east) -| (graph.south);
\end{tikzpicture}
\caption{Implemented data paths. Solid arrows show direct collection and materialization; dashed arrows show optional complete-stream command correlation. Source artifacts remain separate. Display-only path contraction belongs to the viewer, not this batch correlation stage.}
\label{fig:arch}
\end{figure*}
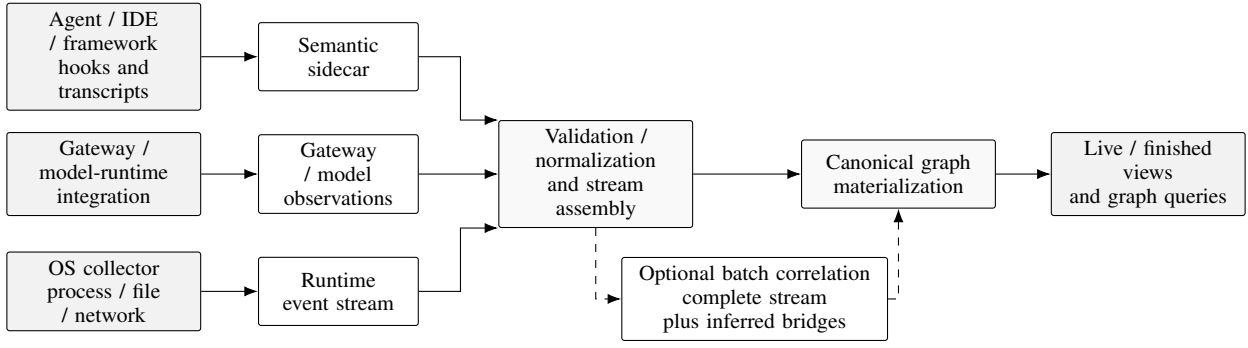

Figure~\ref{fig:arch} shows the data paths. Collection artifacts, graph materialization, and the viewer are local by default; an observed workload can still call remote providers. Provider adapters emit semantic sidecars and runtime collectors independently emit machine observations. Validated and normalized streams can be materialized directly. The optional command correlator consumes a complete event stream and writes a separate correlated stream containing the original observations plus derived bridges, ordered and resequenced for that output. Earlier source files are not rewritten. Correlation is not a mandatory stage of every live or finished run.

\subsection{Semantic collection}
Provider-level evidence can include prompts, responses, tool declarations and results, model identifiers, session identifiers, sub-agent routing, and lifecycle events. Specialized integrations cover Claude Code, OpenAI Codex, Google Antigravity, Cursor, OpenCode, and local or proxied model paths. A framework-adapter interface additionally supplies explicit observation boundaries for CAMEL, AutoGen, and MetaGPT. These integrations expose different subsets of evidence; wrapping an arbitrary framework process alone does not recover its internal agents, messages, or tool arguments.

When an integration supplies content and its capture policy permits storage, \system can preserve the value in a local SHA-256 content-addressed store and place a reference in the event stream. ``Complete from source'' means complete preservation of the supplied representation, not access to hidden model state or remote provider internals. Parsed JSON may be stored in canonicalized form rather than as original transport bytes. Framework content capture is policy-controlled and defaults to metadata-only; policies of other integrations must be checked separately.

\subsection{Runtime collection}
The portable backend combines periodic process/network snapshots with filesystem monitoring on Linux, macOS, and Windows. It follows an observed workload lineage rather than claiming complete OS-wide capture. A process or socket whose lifetime falls between observations can be missed, and scheduling delay can enlarge the effective blind window beyond the configured interval. Filesystem notifications are session-correlated, explicitly non-causal observations: they do not identify the process that performed a change. Missing events are not proof of non-occurrence.

On Linux, a \texttt{strace} backend follows a traced lineage and attributes supported file and network syscalls to processes. Activity outside that lineage, unsupported syscall patterns, and permission restrictions remain outside its scope. File-open relations describe successful opens and access flags, not the bytes subsequently read or written. Failed or pending connection attempts are distinguished from successful connect syscalls. Neither backend provides byte-level information-flow tracking. These differences expose a fidelity/cost trade-off rather than one platform-independent ``captured'' guarantee.

\subsection{Conservative command-to-process correlation}
The optional correlator recognizes a tool-call entity declaring a command entity. Its restricted parser rejects supported categories of shell-control syntax and built-ins when no standalone executable identity can be extracted; it is not a complete shell interpreter or a proof that all expansion constructs have been excluded. For each declaration, the search begins at its timestamp and ends at the earliest applicable bound: the configured maximum window (3 seconds by default), the first recorded success/failure for that tool call, or the immediately following declaration when that declaration has a strictly later timestamp. Both time endpoints are included.

Candidates come from process-start and process-exec observations. Compatibility can be established by executable metadata, command-line head, or an exact nonempty argument tail of matching length. Executable matching accepts a normalized basename match, with resolved-path comparison as another route; it does not require different absolute paths to denote the same binary. Argument-tail matching can admit a process even when its executable head differs. These are intentionally described as heuristics, not exact executable identity tests.

Evidence is accumulated per process identity. If exactly one process remains, the emitted bridge records the method, supporting event IDs, temporal delta, and a heuristic score between 0.80 and 0.95 according to the evidence category. There is no score threshold and no highest-score tie-break between processes. The bridge records \texttt{inferred=true} and \texttt{causal=false}. Multiple candidates, no candidates, and unsupported declarations are separately counted as skip outcomes; the correlator does not serialize every abstained candidate set as a graph node.

\begin{algorithm}[t]
\caption{Optional command-to-process correlation.}
\label{alg:corr}
\footnotesize
\begin{algorithmic}[1]
\Require command declaration $s$ and complete ordered event stream $E$
\Ensure inferred non-causal bridge $b$, or abstention $\bot$
\State $q \gets \Call{ParseDeclaration}{s}$
\If{$q$ is unsupported}
    \State \Return $\bot$
\EndIf
\State $W \gets \Call{BoundWindow}{s,E}$ \Comment{tool result / next declaration / maximum window}
\State $support \gets \{\}$ \Comment{map: process identity $\mapsto$ supporting events}
\ForAll{$e \in \Call{ProcessEvents}{E}$ with $\tau(e) \in W$}
    \If{$\Call{Compatible}{q,e}$}
        \State $p \gets \Call{ProcessId}{e}$
        \State \Call{AccumulateSupport}{$support,p,e$}
    \EndIf
\EndFor
\State $K \gets \Call{Keys}{support}$
\If{$|K| \neq 1$}
    \State \Return $\bot$ \Comment{a score does not resolve ambiguity}
\EndIf
\State $r \gets$ the unique process in $K$
\State $t \gets \Call{ToolCall}{s}$
\State $b \gets \Call{EmitBridge}{t,r}$
\State \Call{RecordSupportAndScore}{$b,support[r]$}
\State \Call{Annotate}{$b$, \texttt{inferred=true}, \texttt{causal=false}}
\State \Return $b$
\end{algorithmic}
\end{algorithm}

This policy cannot eliminate false attribution. Identically named binaries, equal argument tails, parser limitations, window truncation, and missed runtime events can produce false uniqueness. A process reused by several logical actions also need not generate a new process-start event for each action. These cases motivate both abstention accounting and independent negative controls in the evaluation protocol.

\subsection{Exact identity versus causality}
A shared request identifier can link observations across integration points without proving that one event caused another. When a gateway or model-runtime integration supplies such an identifier, the implementation may record \texttt{identity\_exact=true} with \texttt{inferred=false} and \texttt{causal=false}. Provider session IDs and tool-call IDs are logical names; string equality does not make them process IDs. Exact identity is therefore not a substitute for the command correlator and does not place a remote service process into $C_{os}$.

\subsection{Graph materialization and querying}
Each entity collects display attributes, timestamps, event types, and separate counts of logical occurrences and raw evidence. Each relation retains its supporting observations. For causality, the builder aggregates the set of \emph{explicit Boolean flags}: only true yields true, only false yields false, and both or neither yield an unknown value. An omitted flag is ignored by this Boolean set aggregation. Consequently, one true flag plus an unannotated observation can still yield true; the output must not be described as a proof that every support event was causal. Original records and attribution sets remain necessary for more stringent provenance analysis.

Queries filter by node type, relation, backend, or causal status. Directed path queries enumerate bounded simple paths because repeated interactions can introduce cycles. Filtering produces a derived view, not a rewrite of the source graph, and path existence alone does not establish temporal information flow.

\subsection{Live and finished views}
Let $E_t$ be the ordered events incorporated by time $t$. The live view is a projection $P(M(E_t))$ together with fold, selection, and viewport state. Incremental state may omit event-ID lists, apply payload limits, or contain provisional process references. At completion, terminal semantic synchronization and final stream assembly precede the finished graph. The completed-live path installs that final graph, which also supplies the standalone viewer.

Shared graph and conversation machinery makes parity testable, but does not imply equal coverage for every provider/OS pair or equality of every pre-completion cache field. Validation must compare the same finalized stage, identity-resolution policy, and projection settings, separately checking support metadata, content references, and visible conversation history.

\subsection{Semantics-aware visualization}
A root agent, sub-agent, shared model, tool, process tree, file, and external endpoint play different structural roles. The viewer uses layered-layout ideas~\cite{sugiyama1981hierarchical,gansner1993directed,brandes2002horizontal} to organize those roles; dynamic stability remains a task-dependent design choice~\cite{archambault2011dynamic,archambault2013mentalmap}. Neither coordinate optimization nor compactness is evidence of improved analyst performance.

The presentation graph need not be identical to the evidence graph. Hiding intermediate detail can contract a supported path into a display bridge. The implementation marks such bridges \texttt{viewer\_only}, inferred, and non-causal, and records their underlying support paths; ambiguous ownership retains context rather than choosing an agent. These display relations must not be counted as new native observations or confused with the optional batch correlator's output. Layout changes coordinates and routes; projection changes which evidence is displayed. Both remain downstream of the unchanged source graph.

\section{Implementation}
\system is implemented as a Python 3.10+ local command-line and web-viewer system. The implementation described here is release 0.8.25. The package can wrap an arbitrary local command for OS observation or activate provider-specific recorders when stronger semantic integration is available. A run directory can retain raw runtime events, semantic events, correlated events, materialized graphs, conversation views, content-addressed payloads, and a local integrity manifest.

\subsection{Provider-neutral event boundary}
Provider adapters are intentionally upstream of the canonical graph. Their responsibility is to translate exposed lifecycle events and content into a stable event vocabulary while preserving provider identifiers as attributes. This allows downstream graph logic to remain independent of whether the source is a hook, rollout transcript, IDE plugin, gateway event, model-runtime event, or local process collector. It also makes missing semantic coverage explicit: unsupported provider internals remain absent rather than reconstructed from the final answer.

\subsection{Failure and cleanup semantics}
Observability software changes the system it observes and must therefore define failure ownership. The portable collector distinguishes terminal states such as available, degraded, unavailable, not sampled, and not requested for network collection. If a managed collector fails after launching a workload it owns, cleanup is responsible for terminating that owned workload instead of leaving an accidental child behind. Conversely, a descendant that legitimately outlives the root command is not assigned a fabricated exit event merely to close the graph. These distinctions are relevant to both runtime correctness and evaluation under interruption.

\subsection{Privacy and integrity boundary}
Local-first storage reduces default disclosure but does not make captured data harmless. Full-fidelity provider payloads can contain prompts, source code, secrets, tool outputs, file contents, and model responses. The content store is therefore part of the sensitive run artifact. A local integrity manifest can detect changes relative to a recorded manifest, but if both evidence and manifest remain in the same writable trust boundary, this is not an adversary-resistant trusted logging mechanism. The system is an observability tool, not a sandbox or trusted execution environment.

\section{Evaluation Methodology}\label{sec:protocol}
We evaluate \system v0.8.25. For the arXiv v1 evaluation we keep only measurements needed to support the paper's central system claims: targeted negative controls for the optional command-to-process heuristic, direct measurements of the portable collector's sampling boundary, fixed-work runtime-cost measurements, scoped graph-materialization measurements, and one real framework/model/runtime integration. Additional diagnostic fixtures exist in the development tree but are not treated as primary evidence in the manuscript.

\subsection{Targeted attribution negative controls}
The command correlator is not presented as a learned or state-of-the-art matching algorithm. We therefore retain only sanity checks that test its stated contract. Controlled event streams cover two cases that matter directly to the design: \emph{observed ambiguity}, where multiple compatible processes are present and the correlator should abstain, and \emph{false uniqueness}, where the true process is missing and one compatible impostor remains. A forced-choice variant reuses the same parser, time bounds, and candidate construction but chooses the nearest compatible process when several remain. A simpler executable-path-singleton policy provides a minimal baseline. These fixtures use synthetic ground truth generated independently of the correlator output; they are not a second live observer of real process execution.

\subsection{Portable-collector blind window}
The portable backend is sampling based, so its observation boundary must be measured directly. A tracked root workload waits for a randomized phase, launches a short-lived child, and opens a real TCP connection to a harness-owned loopback server for the same phase window. An independent receipt records the child PID, requested child sleep, parent-observed launch-to-wait duration, endpoint port, and connected-socket duration. A process hit requires a periodic \texttt{SPAWNED} event for that oracle child PID; the synchronously recorded root \texttt{LAUNCHED} event cannot count. A socket hit requires the collector's \texttt{CONNECTED\_TO} event for the oracle endpoint.

Requested child sleeps are swept over $\{5,10,25,50,100,250,500,1000\}$~ms. Requested collector intervals are $\{20,25,50,100,250,500\}$~ms; 20~ms is the minimum because v0.8.25 clamps smaller configured intervals to 20~ms. Each cell uses 30 deterministic phase-randomized repetitions. We report hits out of 30, empirical detection probability, and Wilson 95\% intervals. Requested sleep is not treated as measured process lifetime: the parent-observed duration is retained separately. Process and socket hit labels are derived from the RuntimeCollector stream and the oracle receipt rather than from a second collector.

\subsection{Fixed-work paired overhead}
Performance uses fixed completed work rather than time-budget loops. The micro workload performs 64 file-write/read/hash iterations and a fixed child-process schedule; medium performs 4,000 iterations; long performs 20,000. The off and portable conditions within a repetition execute the same work specification and must produce receipts with equal completed iteration counts, child counts, and work digests. A primary row fails closed when this equal-work contract is not satisfied.

For each repetition, the order of off and portable is deterministically randomized. Micro and medium use 30 pairs and long uses 10. Startup and shutdown remain inside elapsed time. We report standard medians and distribution summaries for elapsed time, whole measured process-tree peak RSS, artifact bytes, and event counts. Memory is therefore not described as collector-only RSS or proportional-set size. Paired relative overhead is $T_{portable}/T_{off}-1$ for each pair; the primary summary reports the median paired ratio with a deterministic 5,000-resample bootstrap 95\% interval. If \texttt{strace} is installed on the experiment host, it is retained as an optional characterization arm rather than pooled into the primary off/portable comparison.

\subsection{Scoped graph materialization}
A synthetic canonical-event generator is applied at 100, 1,000, 5,000, 10,000, and 50,000 events, with five repetitions per scale. Unique resources are explicitly capped at 10,000, so the 50,000-event point contains approximately 10,001 nodes rather than 50,000 nodes. The accumulator uses \texttt{retain\_event\_ids=false}. Timers separately measure event application, \texttt{GraphAccumulator.to\_dict()}, and JSON encoding of the resulting snapshot; the benchmark does not include disk transfer, network transfer, or browser layout/rendering time.

\subsection{Reproducibility boundary}\label{sec:repro}
The measurements use \system v0.8.25 on Ubuntu~24.04 with Python~3.12.11. Source code is available at \url{https://github.com/Irish-kw/ExecWeave}. The numbers in Section~\ref{sec:results} are author-run measurements on that release. We do not backfill missing cells from older releases.

The broader evaluation remains future work rather than being inferred from this Linux campaign. Matched Windows/macOS runs, additional provider/framework executions with independent ground truth, direct product comparisons, stronger whole-system provenance backends under the same workload, and a controlled human debugging study are intentionally outside the arXiv v1 scope.

\subsection{Real AutoGen--Ollama integration}
To verify that the evidence contract survives a real framework/model execution, we run the v0.8.25 AutoGen acceptance workload with AutoGen~0.7.5 and a local Ollama \texttt{qwen2.5:0.5b} model on Linux. The portable \texttt{RuntimeCollector} wraps the framework process with a 20-ms effective sampling interval while the framework adapter writes its semantic sidecar.

The acceptance contract is evidence based. The real AutoGen group chat must complete with model responses and no model failure; the portable stream must contain the framework root process and a real \texttt{CONNECTED\_TO} observation for the configured Ollama endpoint; semantic process references must resolve against the observed process during the \texttt{merge\_semantic\_sidecar} step; and the merged stream must materialize into a graph containing both agent and process identities. This uses the implementation's own AutoGen acceptance path rather than a replacement mock.

This is a single-case end-to-end validation, not a provider benchmark. It establishes that one concrete AutoGen/Ollama execution traverses semantic capture, independent portable process/network observation, process-reference resolution, and canonical graph materialization. It does not establish cross-framework, cross-provider, cross-model, or cross-OS parity.

\section{Empirical Evaluation Results}\label{sec:results}
The measurements below use Ubuntu~24.04, Python~3.12.11, and \system v0.8.25. The AutoGen--Ollama integration is a separate single-run check. The manuscript reports only the measurements needed to support the core system claims.

\subsection{Targeted attribution sanity checks}
The optional correlator behaves as specified when ambiguity is visible: with multiple compatible observed processes it abstains, whereas a forced-nearest policy introduces false associations by choosing among those candidates. The more important negative control is false uniqueness. When the true process is deliberately absent and one compatible impostor remains, the singleton policy can emit the wrong bridge. A simpler executable-path-singleton policy also matches the full correlator on the controlled fixture suite. We therefore use these fixtures only to validate the abstention contract and expose its boundary; they do not support a claim that the full heuristic matcher is more accurate than simpler alternatives.

\subsection{Portable sampling boundary}
The direct RuntimeCollector experiment exposes the expected blind window. Table~\ref{tab:sampling-results} shows representative cells from the complete $8\times6$ sweep. Each cell has 30 phase-randomized repetitions. Process and socket hit labels are taken from the native event stream together with the oracle child PID and endpoint.

\begin{table*}[t]
\centering
\caption{Representative RuntimeCollector sampling results ($n=30$ per cell). Requested sleep is distinct from the measured parent-observed child duration. P/S = process hits / socket hits.}
\label{tab:sampling-results}
\footnotesize
\setlength{\tabcolsep}{3.2pt}
\renewcommand{\arraystretch}{1.06}
\begin{tabularx}{\textwidth}{@{}
>{\hsize=0.85\hsize\linewidth=\hsize\centering\arraybackslash}X
>{\hsize=1.35\hsize\linewidth=\hsize\centering\arraybackslash}X
>{\hsize=0.85\hsize\linewidth=\hsize\centering\arraybackslash}X
>{\hsize=0.75\hsize\linewidth=\hsize\centering\arraybackslash}X
>{\hsize=1.05\hsize\linewidth=\hsize\centering\arraybackslash}X
>{\hsize=1.15\hsize\linewidth=\hsize\centering\arraybackslash}X@{}}
\toprule
Requested sleep & Median observed child duration & Poll interval & P/S hits & Process probability & Process Wilson 95\% CI \\
\midrule
5 ms & 16.38 ms & 20 ms & 18/19 & 0.600 & $[0.423,0.754]$ \\
5 ms & 16.21 ms & 500 ms & 0/0 & 0.000 & $[0.000,0.114]$ \\
25 ms & 36.16 ms & 50 ms & 17/20 & 0.567 & $[0.392,0.726]$ \\
50 ms & 61.40 ms & 100 ms & 21/21 & 0.700 & $[0.521,0.833]$ \\
250 ms & 262.64 ms & 500 ms & 13/13 & 0.433 & $[0.274,0.608]$ \\
500 ms & 512.77 ms & 500 ms & 30/30 & 1.000 & $[0.886,1.000]$ \\
\bottomrule
\end{tabularx}
\end{table*}

For a requested 5-ms child, whose median parent-observed duration is approximately 16~ms, process detection falls from 18/30 at a 20-ms interval to 0/30 at 500~ms. A requested 50-ms child is observed in 21/30 runs at a 100-ms interval; a requested 250-ms child is observed in 13/30 runs at 500~ms. The supported conclusion is empirical rather than a deterministic threshold: on this host, detection degrades as the polling interval grows relative to the observable lifetime, with scheduling and startup overhead affecting individual cells.

\subsection{Fixed-work overhead}
Table~\ref{tab:overhead-results} reports paired off-versus-portable measurements in which both members complete the same iterations, child-process count, and work digest; backend order is randomized within each pair. Portable observation adds a median 408.8\% elapsed time on the micro workload, 85.2\% on medium, and 62.3\% on long. The bootstrap intervals remain well above zero for all three workloads.

\begin{table*}[t]
\centering
\caption{Fixed-work portable overhead. The primary statistic is the within-pair relative elapsed-time increase. RSS is peak whole-process-tree RSS, not collector-only memory or PSS.}
\label{tab:overhead-results}
\footnotesize
\setlength{\tabcolsep}{2.8pt}
\renewcommand{\arraystretch}{1.06}
\begin{tabularx}{\textwidth}{@{}
>{\hsize=0.65\hsize\linewidth=\hsize\centering\arraybackslash}X
>{\hsize=0.55\hsize\linewidth=\hsize\centering\arraybackslash}X
>{\hsize=1.20\hsize\linewidth=\hsize\centering\arraybackslash}X
>{\hsize=1.05\hsize\linewidth=\hsize\centering\arraybackslash}X
>{\hsize=1.20\hsize\linewidth=\hsize\centering\arraybackslash}X
>{\hsize=1.00\hsize\linewidth=\hsize\centering\arraybackslash}X
>{\hsize=1.35\hsize\linewidth=\hsize\centering\arraybackslash}X@{}}
\toprule
Workload & Pairs & Paired median overhead & P25--P75 & Bootstrap 95\% CI & Portable wall median & Portable RSS / artifact \\
\midrule
Micro & 30 & +408.8\% & +402.6--+420.6\% & +404.8--+417.1\% & 0.470 s & 61.7 MiB / 1.00 MiB \\
Medium & 30 & +85.2\% & +74.2--+98.2\% & +78.4--+96.2\% & 2.390 s & 70.0 MiB / 5.53 MiB \\
Long & 10 & +62.3\% & +52.6--+70.3\% & +51.8--+71.2\% & 6.692 s & 104.1 MiB / 12.43 MiB \\
\bottomrule
\end{tabularx}
\end{table*}

The corresponding unobserved median wall times are 0.0918, 1.2853, and 4.1217~s. These measurements do not support an unqualified ``low-overhead'' characterization. Instead, they show substantial fixed and sustained cost for these file- and child-process-heavy fixtures, together with increasing retained artifact volume and process-tree memory. The syscall-oriented Linux backend is retained only as descriptive characterization in the artifact and is not used as a controlled backend ranking in the manuscript.

\subsection{Scoped graph materialization}
Table~\ref{tab:scale-results} separates event application from graph snapshot materialization and JSON encoding. Unique resources are capped at 10,000; therefore the 50,000-event point contains 10,001 nodes and 10,000 edges, not 50,000 nodes.

\begin{table*}[t]
\centering
\caption{Graph-materialization medians over five repetitions. ``Snapshot'' is \texttt{to\_dict()} and ``JSON'' is encoding of that snapshot. Browser rendering, disk transfer, and network transfer are excluded.}
\label{tab:scale-results}
\footnotesize
\setlength{\tabcolsep}{2.8pt}
\renewcommand{\arraystretch}{1.06}
\begin{tabularx}{\textwidth}{@{}
>{\hsize=0.90\hsize\linewidth=\hsize\centering\arraybackslash}X
>{\hsize=1.00\hsize\linewidth=\hsize\centering\arraybackslash}X
>{\hsize=0.90\hsize\linewidth=\hsize\centering\arraybackslash}X
>{\hsize=0.90\hsize\linewidth=\hsize\centering\arraybackslash}X
>{\hsize=1.05\hsize\linewidth=\hsize\centering\arraybackslash}X
>{\hsize=1.05\hsize\linewidth=\hsize\centering\arraybackslash}X
>{\hsize=0.95\hsize\linewidth=\hsize\centering\arraybackslash}X
>{\hsize=1.25\hsize\linewidth=\hsize\centering\arraybackslash}X@{}}
\toprule
Events & Resources & Nodes & Edges & Apply & Snapshot & JSON & Snapshot bytes \\
\midrule
100 & 100 & 101 & 100 & 1.36 ms & 1.47 ms & 0.63 ms & 95,214 \\
1,000 & 1,000 & 1,001 & 1,000 & 12.92 ms & 14.70 ms & 6.29 ms & 942,122 \\
5,000 & 5,000 & 5,001 & 5,000 & 76.36 ms & 91.47 ms & 30.18 ms & 4,730,122 \\
10,000 & 10,000 & 10,001 & 10,000 & 162.17 ms & 210.84 ms & 59.56 ms & 9,465,130 \\
50,000 & 10,000 & 10,001 & 10,000 & 579.19 ms & 190.51 ms & 60.14 ms & 9,476,236 \\
\bottomrule
\end{tabularx}
\end{table*}

At 50,000 events the median application time is 0.579~s, approximately 86.3 thousand events/s on this runner, while snapshot plus JSON encoding takes approximately 0.251~s. Because the graph topology is capped from 10,000 resources onward, this result characterizes repeated-event ingestion and serialization for a bounded unique-resource graph; it does not establish 50,000-node scalability or browser-interaction performance.

\subsection{Real AutoGen--Ollama integration}
The real integration completes four model requests and four model responses with zero model failures. RuntimeCollector observes the framework root process and a real connection to the Ollama endpoint. All three semantic process references resolve to the observed framework process, and the final materialized graph contains 60 nodes and 61 edges, including three agent-to-process \texttt{CORRELATED\_WITH\_PROCESS} relations.

This result closes a narrow but important evidence gap: one real framework, one real local model runtime, independent portable OS observation, semantic/runtime process-reference resolution, and canonical graph materialization all traverse the frozen implementation end to end. It does not establish equivalent fidelity across AutoGen versions, other frameworks, hosted providers, remote MCP servers, or other operating systems.

\section{Discussion}
\subsection{What the evaluation establishes}
The strongest evidence is end-to-end and boundary oriented rather than a claim of superior matching accuracy. The AutoGen--Ollama run verifies one concrete semantic/runtime integration path. The portable-collector sweep quantifies when short-lived process and socket evidence can disappear. Fixed-work pairs expose the runtime and storage cost of observation, and the graph benchmark characterizes materialization and serialization separately. The targeted attribution checks support only the intended epistemic rule: abstain when observed candidates are ambiguous, and do not mistake a singleton observed candidate for causal proof when evidence may be missing.

\subsection{Observation fidelity and performance are material limitations}
The portable blind window is not merely hypothetical. Absence of an event becomes weaker evidence as object lifetime approaches or falls below the effective polling interval, and a missed true process can create false uniqueness at the attribution layer. The fixed-work results are an equally important negative result: portable observation adds substantial elapsed time on all three workloads measured here, while memory and artifact volume also grow with observation. These costs may be acceptable in debugging or forensic settings but cannot be described as generally low without a broader deployment study.

\subsection{The graph is a claim surface, not merely a view}
An execution graph is persuasive because edges visually imply relationships. That makes graph construction a claim boundary. If a system draws a tool directly to a process, users naturally interpret the line as evidence that the tool caused or at least executed through that process. \system therefore treats every cross-layer edge as something that must be explainable through support metadata. Explainability does not itself guarantee causal correctness when observations are incomplete, which is why inferred and non-causal labels remain part of the graph contract.

\subsection{Relationship to AgentSight and provenance systems}
AgentSight already demonstrates that agent semantics and kernel-level system effects can be combined for AI-agent observability~\cite{zheng2025agentsight}. \system therefore does not claim the general semantic--OS combination as its novelty. Its narrower contribution is an evidence-typed artifact and graph contract spanning heterogeneous integration sources, explicit non-causal marking of heuristic bridges, inspectable supporting observations, and a portable observation path in addition to a stronger Linux backend. The present evaluation does not establish superiority over AgentSight.

Whole-system provenance systems provide stronger causal capture boundaries than \system's portable collector in several respects~\cite{pohly2012hifi,gehani2012spade,bates2015lpm,pasquier2017camflow}. The specialization here is the preservation of agent, tool, model, provider, and runtime identities alongside OS entities, not a replacement for kernel provenance or byte-level information-flow tracking.

\subsection{Compatibility with OpenTelemetry}
The model is not an argument against OpenTelemetry. OpenTelemetry is a natural transport and semantic source for instrumented agent and service operations~\cite{otel2026}. The key design question is whether evidence was emitted by the application, independently observed at runtime, or derived by correlation. A future interoperability layer could ingest OTel spans as $C_{sem}$ or $C_{gw}$ events while preserving independent $C_{os}$ observations. Cross-layer links could then use propagated trace/span context as strong identity support when available and fall back to conservative runtime evidence otherwise. Ingesting an OpenTelemetry span into $C_{sem}$ or $C_{gw}$ does not move it into $C_{os}$.

\subsection{Cross-platform and remote-execution boundaries}
A portable collector cannot truthfully promise kernel-equivalent fidelity across Windows, macOS, and Linux using only common user-space APIs. \system therefore favors a common canonical schema with backend-specific evidence strength over pretending that all platforms expose identical telemetry. The measured campaign is Linux-only; Windows/macOS parity remains unmeasured.

Local OS observation also stops at the local machine. A call to a hosted model can expose local client/network evidence and provider-supplied semantics, but it cannot reveal the provider's remote process tree unless the provider supplies additional telemetry. The same applies to a remote Model Context Protocol (MCP) server: local evidence can establish the client call and network boundary, not the tool's internal remote behavior. File/network co-occurrence similarly does not prove byte-level flow, exfiltration, or malicious intent. Evidence strength and behavioral severity remain separate dimensions.

\section{Threats to Validity}
\subsection{Construct validity}
The targeted attribution checks evaluate a deliberately narrow command-declaration--to--observed-process relation rather than causal responsibility. Their expected outcomes are generated by the fixture constructor independently of the correlator output, but they remain synthetic fixture ground truth rather than a second live observer of process execution. The result supports the stated abstention behavior and exposes false uniqueness under missing evidence; it does not establish that the full heuristic matcher outperforms simpler policies.

The portable-collector experiment measures the actual \texttt{RuntimeCollector}, but requested child sleep is not identical to OS-visible process lifetime. The independent receipt therefore records both the requested sleep and the parent-observed launch-to-wait duration. The loopback socket is a controlled local connection rather than representative Internet traffic. The graph-materialization benchmark caps unique resources at 10,000; its 50,000-event point tests repeated-event application and serialization, not a 50,000-node graph.

The AutoGen--Ollama validation is intentionally a single end-to-end case. It demonstrates that one framework/model execution traverses semantic capture, independent runtime observation, process-reference resolution, and graph materialization, but it is not a provider benchmark or evidence of broad framework parity.

\subsection{Internal validity}
The attribution negative controls are deterministic by construction, so their principal risk is fixture logic rather than model randomness. The fixture suite includes raw rows, duplicate-key checks, and negative controls, but an error shared by a fixture generator and its oracle could still survive unless independently inspected.

Sampling is timing-sensitive. The portable-collector sweep randomizes phase deterministically within each lifetime/interval cell and reports Wilson intervals rather than treating an isolated hit count as a deterministic law. Process and socket detection labels are derived from the native RuntimeCollector stream and the workload's child-PID/endpoint receipt.

The fixed-work overhead protocol verifies equal completed iteration counts, child counts, and work digests between the off and portable members of each pair. Backend order is randomized within each pair to reduce monotone warm-cache or host-drift effects. Nevertheless, elapsed time, scheduler behavior, filesystem caching, process-tree RSS, and artifact growth remain host dependent. Startup and finalization are deliberately included. The measured RSS is peak whole-process-tree RSS; it is neither collector-only resident memory nor proportional-set size.

\subsection{External validity}
The principal measurements are Linux-only. They do not establish empirical parity across Windows and macOS, every provider integration, remote MCP internals, hosted observability products, browser agents, Kubernetes workloads, serverless tool execution, or remote sandboxes. The portable implementation supports multiple operating systems, but implementation support and measured cross-platform equivalence are separate claims.

The real integration surface has the same boundary. One AutoGen version with one local Ollama model does not imply equivalent coverage for other versions, frameworks, providers, or models. Remote model and tool internals remain outside local OS visibility unless separately instrumented. Product comparisons must likewise distinguish documented native/default collection, documented extensibility, and experimenter-supplied custom spans; absence from documentation is not proof that a general telemetry platform could never represent an externally supplied OS event.

\subsection{Conclusion validity}
The reported measurements are descriptive results from controlled repeated workloads, not independent samples from a population of real agent executions. We therefore report underlying counts and uncertainty where appropriate rather than derive significance claims from pooled event rows. For the sampling sweep, $n=30$ per cell supports uncertainty intervals but not fine-grained claims from small differences between neighboring cells. For overhead, the primary statistic is the within-pair ratio rather than a ratio of aggregate medians.

Human-debugging effectiveness is not measured. Claims about analyst accuracy, time-to-root-cause, confidence calibration, visual comprehensibility, or subjective workload require a separately powered human-subject study. Performance results must likewise be interpreted jointly with memory and artifact growth rather than compressed into an unqualified ``low overhead'' label.

\section{Reproducibility, Ethics, and Data Handling}
The implementation evaluated in this manuscript is ExecWeave v0.8.25. Source code is available at \url{https://github.com/Irish-kw/ExecWeave}. The measurements reported in Section~\ref{sec:results} are author-run results on that release. Unmeasured cells are left unmeasured; they are not filled in from older releases.

Observability studies can capture sensitive prompts, source code, authentication material, file contents, and network destinations. The controlled fixtures use no participant histories, and the framework/model acceptance run uses experiment-owned prompts and a local endpoint. Any future human-subject evaluation should minimize captured content, define retention before collection, and use an appropriate review process. Public traces should be synthetic or independently reviewed for credentials and private material.

Attribution also has an ethical dimension. A graph naming an agent or process can be interpreted as assigning responsibility. \system therefore exposes provenance, inference, and causal status instead of presenting heuristic association as forensic certainty. The false-uniqueness negative demonstrates why this qualification is substantive: abstaining on observed ambiguity does not protect against an unobserved true process plus one observed impostor.

\section{Future Research Directions}
Several technical extensions follow directly from the measured limitations. Propagated request/process identities could replace heuristic matching when available, but require adoption across agent, tool, gateway, and runtime boundaries. Learned correlation might improve recall, yet any learned policy would need calibrated abstention, inspectable supporting features, and explicit missing-observation tests. Stronger whole-system provenance collectors could improve Linux fidelity while preserving the same agent-oriented evidence contract, and distributed agents require provenance across machines and trust domains rather than a single local process tree.

The next empirical priorities are to repeat the portable sampling and fixed-work protocols on Windows and macOS, expand real provider/framework end-to-end runs with independent ground truth, compare representative observability systems under matched instrumentation conditions, scale unique graph entities as well as repeated events, and conduct a controlled human study of evidence-backed debugging tasks. These are deliberately left for later revisions rather than being inferred from the arXiv v1 campaign.

\section{Conclusion}
Agent traces and operating-system telemetry answer different questions. Joining them is therefore not merely a visualization operation: every cross-layer relation carries an evidentiary claim. \system makes that claim explicit by preserving semantic and runtime observations as distinguishable evidence, materializing them into one typed graph, and marking heuristic command-to-process bridges as inferred and non-causal.

The arXiv v1 evaluation focuses on the evidence needed to support that design rather than treating every internal fixture as a headline benchmark. A real AutoGen--Ollama execution validates one end-to-end semantic/runtime path; targeted attribution negative controls show why observed ambiguity should produce abstention and why missing evidence can still create false uniqueness; direct RuntimeCollector measurements quantify the portable blind window; fixed-work pairs measure observation cost; and graph-materialization measurements separate event application from snapshot and JSON encoding.

The bounded conclusion is methodological as well as architectural: cross-layer observability is most useful when a graph preserves where each observation came from, distinguishes identity from causality, retains support for derived relations, and permits uncertainty to remain visible. The experiments do not establish universal platform parity, product superiority, or human-debugging benefit, and the paper does not claim otherwise.

\IEEEtriggeratref{30}
\bibliographystyle{IEEEtran}
\bibliography{references}

\end{document}